\documentclass{article}

\usepackage{amsmath}



\newcounter{prop}[section]
\newcounter{theor}[section]
\newcounter{defin}[section]
\newcounter{theor_append}

\newenvironment{defin}
{ \refstepcounter{defin}
 {\ \newline
\bf Definition \arabic{section}.\arabic{defin}.}\em} {}

\begin{document}

\title{ New Approach for Stochastic Quantum Processes,
Their Manipulation and Control }
\author{Ashot S. Gevorkyan }

\date{Institute for Informatics and Automation Problems, NAS RA,\\
           Ind. 375014, Parujr Sevak str., H. 1, Yerevan, Armenia,\\
               g\_ashot@sci.am \\}

\maketitle

\begin{abstract}

The dissipation and decoherence (for example, the effects of noise
in quantum computations), interaction with thermostat or in
general with physical vacuum, measurement and many other
complicated problems of open quantum systems are a consequence of
interaction of quantum system with the environment. These problems
are described mathematically in terms of complex probabilistic
process (CPP). Particularly, treating the environment as a
Markovian process we derive an Langevin-Schr\"{o}dinger type
stochastic differential equation (SDE)  for describing the quantum
system interacting with environment. For the $1D$ randomly quantum
harmonic oscillator (QHO) model L-Sh SDE is a solution in the form
of orthogonal CPP. On the basis of orthogonal CPP the stochastic
density matrix (SDM) method is developed and in its framework
relaxation processes in the uncountable dimension closed system of
"QHO+environment" is investigated. With the help of SDM method the
thermodynamical potentials, like nonequilibrium entropy and the
energy of ground state are exactly constructed. The
 dispersion for different operators are calculated.
In particular, the expression for uncertain relations depending on
parameter of interaction with environment is obtained. The Weyl
transformation for stochastic operators is specified. Ground state
Winger function is developed in detail.
\end{abstract}


\section*{Introduction}
\label{sec-0}

Recently a great number of papers \cite{proc} concerning the
"quantum chaos", i.e. with the quantum analogues of classical
systems possessing the dynamic chaos features, have been
published. The investigations are conducted along different
directions, such as analysis of distribution of energy levels;
definition and calculation of quantities, which are responsible
for the presence of chaos (corresponding to the classical Lyapunov
exponents and KS-entropy)in the quantum systems; study of
localization and delocalization of wave functions around the
classical orbits; etc. Though in most of cases mentioned above one
is faced with the necessity to describe a quantum system
statistically, so far there was not paid much attention to a
stochastic behavior of the wave function itself.

Many problems of great importance in the field of the
non-relativistic quantum mechanics, such as description of Lamb
shift, spontaneous transitions in atoms, etc., remain unsolved due
to the fact that the concept of environment (which as a rule is
random) has not been considered within the framework of the
standard quantum mechanics. It is obvious that a quantum object
immersed into the thermostat (or more fundamentally the physical
vacuum) is an open system. Various approaches \cite{pot} to the
description of such systems exist, mainly in application to the
problem of continuous measurements. One of them is based on the
consideration of the wave function as a random process, for which
a stochastic differential equation (SDE) is derived. But the
equation is obtained by the method which is extremely difficult
for application even in case of comparatively simple type of
interaction between the system and the environment, so that some
new ideas are needed \cite{gc}-\cite{sinai}. Moreover sometimes it
becomes necessary to consider the wave function as a random
complex process even in closed systems (for example, when a
classical analogue of the quantum system has the features of the
dynamical chaos) \cite{avt1}-\cite{avt3}.

To describe the cases mentioned above, recently was proposed a
radically new mathematical scheme (theory of stochastic quantum
processes (TSQP). This scheme is used to describe the
non-relativistic quantum system, in general case strongly
interacting with the  randomly environment \cite{ash1}. Note, that
the wave function of closed system "quantum object + thermostat"
at that time is described by L-Sch type SDE. On the base of TSQP
was developed stochastic density matrix method permitting
construction of  all the thermodynamic potentials of quantum
subsystem.

In the present paper relaxation processes in a closed uncountable
dimension system ''quantum oscillator and thermostat'' are
investigated  within the framework of non-perturbational  method.

In particular is receipted evolution equation for described
excitations into thermostat. Exact representations are found for
both widening and shift (analogous to the Lamb shift) of the
ground state energy level of the immersed into the thermostat
(physical vacuum) QHO as well as the entropy of an ground state is
calculated.


\section{Formulation of the problem}
\label{sec-1}
 We shall consider the closed system "quantum object
+ thermostat" within the framework of Langevin-Schr\"{o}dinger
type SDE
\begin{equation}
i\partial_{t}\Psi_{stc}=\hat{H} \Psi_{stc}, \label{1.01}
\end{equation}
where 1D evolution operator $ \hat H $ is assumed to be quadratic
over the space variable:
\begin{equation}
\hat H = -\frac {1}{2}\frac {\partial^2}{\partial x^2}+ \frac
{1}{2}\Omega^2(t)x^2.
\label{1.02}
\end{equation}
In expressions (\ref{1.01})-(\ref{1.02}) the frequency {$
\Omega(t) $ is stochastic function of time. Let them have the
form:
\begin{eqnarray}
\Omega^2(t) = \Omega_{0}^2 + \sqrt{2\epsilon}f(t), \label{1.03}
\end{eqnarray}
where $\Omega_{0}=const $ and $ f(t)$ is independent Gaussian
stochastic process with the zero mean and $ \delta - $ shaped
correlation function:
\begin{equation}
<f(t)f(t')> = \delta (t-t'),\qquad <f(t)>=0. \label{1.04}
\end{equation}
Constant $ \epsilon\ $  is characterized  the power of stochastic
force $ f(t) $. The equation  (\ref{1.01})  have asymptotic
solution $ \Psi_{as}(n|x,t) $ in the limit of $t\rightarrow
-\infty $:
\begin{eqnarray}
\Psi_{as}(n|x,t) = e^{-i(n+1/2)\Omega_{0}t}\phi(n|x), \nonumber
\\
\phi(n|x) = \biggl(\frac {1}{2^nn!}\sqrt{\frac
{\Omega_{0}}{\pi}}\,
\biggr)^{1/2}e^{-\Omega_{0}x^2/2}H_n\Bigl(\sqrt{\Omega_{0}}x\Bigr),
\label{1.05}
\end{eqnarray}
where $ \phi(n|x) $ is the wave function of a stationary
oscillator and $ H_n(x) $ is the Hermitian polynomial. The formal
solution of the problem (\ref{1.01})-(\ref{1.05}) may be written
down explicitly for arbitrary $ \Omega(t) $. It has the following
form:

\begin{eqnarray}
\Psi_{stc}(x,t\vert \{ \xi\}) =\frac {1}{\sqrt{r}}exp\Bigl
\{\frac{i}{2}\dot{r}r^{-1}x^2\Bigr\}\chi \Bigl(\frac
{x}{r},\tau\Bigr), \label{1.06}
\end{eqnarray}
where the function $ \chi (y,\tau ) $ satisfies the
Schr\"{o}dinger equation for a harmonic oscillator on the
stochastic space-time $ \{y,\tau\} $ continuum:

\begin{equation}
\label{1.07}
i\frac {\partial \chi }{\partial \tau }=-\frac 1 2 \frac
{\partial^2\chi }{\partial y^2}+\frac {\Omega_{0}^2y^2}{2}\chi,
\end{equation}
where
$$ y= \frac{x}{r}, \quad \xi (t)=r(t)e^{i\gamma (t)},
\quad \tau =\frac{\gamma (t)}{\Omega_{0}},\quad \gamma
(t)=\Omega_0\int\limits_{-\infty}^t\frac{dt'}{r^2(t')}.
$$
The function $ \xi (t) $ is defined from the classical homogenous
equation of motion for the oscillator with the frequency
$\Omega(t)$
\begin{equation}
\ddot \xi +\Omega^2(t)\xi =0. \label{1.08}
\end{equation}

Taking into account (\ref{1.06}) and well known solution of
autonomous  quantum harmonic oscillator \cite{Baz'} for stochastic
complex processes which are described closed strong interacting
system ''quantum object+environment'' we can write following
expression:

$$
\Psi_{stc}(m|x,t; \{ \xi\})=\biggl(\frac {1}{2^nn!}\sqrt{\frac
{\Omega_{0}}{\pi r(t)}}\,\biggr)^{1/2}\times
$$
\begin{equation}
\exp\biggl[-i\Bigl(n+\frac12\Bigr)\Omega_{0}\int\limits_{t_0}^t
\frac{dt'}{r^2(t')}+i\frac{r_t(t)}{2r(t)}x^2-\frac{\Omega_{0}}
{2r^2(t)}x^2\biggr]H_n\Bigl(\sqrt{\Omega_{0}}\frac{x}{r(t)}\Bigr).
\label{1.09}
\end{equation}
The solution of (\ref{1.09}) is a random complex process defined
on the extended space $\Xi= R^1\otimes{R_{\{\xi\}}}$, where $R^1$
is one dimensional euclidian space and $R_{\{\xi\}}$ is the
corresponding functional space. It is easy to show that the
mentioned complex random processes are orthogonal. Taking integral
over the space $R^1$ we get:
\begin{equation}
\int\limits_{-\infty}^{+\infty}\Psi_{stc}(n|x,t; \{ \xi\})
\Psi_{stc}^{\ast}(m|x,t; \{ \xi\})\,dx=\delta_{nm}, \label{1.10}
\end{equation}

 where the symbol $\ast$ means the complex conjugation.
The relation (\ref{1.10}) shows that the closed odd dimensional
system "quantum oscillator+environment" is described in terms of
the full orthogonal basis of quadratically integrable functionals
of the space $L^2$. The last fact is very important for the
further strong mathematical constructions for the statistical
parameters of the system.


\setcounter{equation}{0}
\section{Stochastic density matrix method}
\label{sec-2}
 The quantum system is impossible to isolate from the environment.
This is a principal problem if taking into account that even for
the ideal isolation nevertheless any system is located in the
fundamental physical vacuum. Remember that many important features
of atomic systems, like Lamb shift of energy levels, spontaneous
transitions, etc., are explained by vacuum fluctuations. So, the
processes in quantum systems are to some extent irreversible. For
the investigation of irreversible processes the non-stationary
density matrix representation based on quantum Liouvile equation
\cite{Zubar} is often used. However the application of this
representation has restrictions. It is used for the cases when the
system before the interaction switch on was on thermodynamic
equilibrium state and after the interaction was applied, its
evolution is adiabatic. Below in the frames of considered model
the new approach is used for  the investigation of statistical
properties of irreversible quantum system without any restriction
on the quantities and the rate of interaction change. This method
is based on bilinear form is constructed from orthogonal complex
random processes. Below it is referred as the \emph{stochastic
density matrix method}.

\begin{defin}
The stochastic density matrix is defined by the expression:
\begin{equation}
\label{2.01}
 \rho_{stc}( x,t;\{\xi\} \vert x^{\prime },t^{\prime
};\{\xi^{\prime }\}) = \sum_{m=0}^\infty w_0^{( m) }\rho_{stc}^{(
m) }( x,t;\{ \xi\} \vert x^{\prime },t^{\prime }; \{ \xi^{\prime
}\}) ,
\end{equation}
where $\rho_{stc}^{( m) }( x,t;\{ \xi\} \vert x^{\prime
},t^{\prime }; \{ \xi^{\prime }\})$ is partial stochastic density
matrix and defined by help of bilinear form:
\begin{equation}
\label{2.02}
 \rho_{stc}^{( m)}( x,t;\{ \xi \} \vert x^{\prime },
t^{\prime };\{ \xi^{\prime }\}) = \Psi_{stc}(m|x,t; \{\xi\})
\Psi_{stc}^{\ast}(m| x^{\prime },t^{\prime}; \{ \xi^{\prime }\}).
\end{equation}
 In the expression of (\ref{2.01}) $w_0^{( m)}$ has the meaning
of the initial distribution over quantum states with energies
$E_m=(m+ 1/2) \Omega _{0}$, until the moment when the generator of
random excitations is activated. Integrating (\ref{2.01}) over
euclidian space and taking into account (\ref{1.10}), we obtain
the normalization condition for the weight functions:
\begin{equation}
\label{2.03}
 \sum_{m=0}^\infty w_0^{( m)}=1,\quad w_0^{( m)}\geq 0.
\end{equation}
\end{defin}
Below we'll define the mean values of various operators. Note that
when averaging on extended space $\Xi$ the order of the
integration is important. If the integral is taken first on $R^1$
space then on $R_{\{\xi\}}$, the stochastic density matrix becomes
equal to unity. This means that  in the extended space all
conservations laws are valid, in other words the stochastic matrix
in this space is unitary. Else if we take the integration in the
inverse order, we get another picture. After the integration on
$R_{\{\xi\}}$ the obtained density matrix describes quantum
processes in Euclidean space $R^1$. Its trace is not unity, in
general. This means that the conservation laws are not valid
already. This can be explained by the fact that the system has
been in a strongly non-equilibrium state and after relaxation its
parameters have been significantly changed.

Below we'll be interested in quantum subsystem processes, hence
the integration first on $R_{\{\xi\}}$ and then on $R^1$ is
supposed.

\begin{defin} The expected value of the operator
$\hat A( x,t\vert \{ \xi\})$ in quantum state with the index $m$
is:
\begin{equation}
\label{2.04}
 A_m= \lim\limits_{t\rightarrow +\infty }\biggl\{\frac{1}{N_{m}(t)}
 Sp_x\Bigl[ Sp_{\{ \xi\} }\hat A\rho_{stc}^{( m) }\Bigr]\biggr\},
 \quad N_{m}(t)=Sp_x\Bigl[Sp_{\{\xi\} }\rho _{stc}^{( m)}\Bigr].
\end{equation}
The mean value of the operator $\hat A( x,t\vert \{\xi \}) $ over
the whole ensemble of states will respectively be given by:
\begin{equation}
\label{2.05} A= \lim\limits_{t\rightarrow +\infty }\biggl
\{\frac{1}{N(t)}Sp_x\Bigl[ Sp_{\{ \xi\} }\hat{A}\rho _{stc}\Bigr]
   \biggr\}, \quad
N(t)=Sp_x\Bigl[Sp_{\{\xi\} } \rho_{stc}\Bigr].
\end{equation}
\end{defin}
The operation $ Sp_{\{\xi\}} $ in (\ref{2.03}) and (\ref{2.04}) is
defined by functional integral representation
\begin{equation}
\label{2.06}
 Sp_{\{\xi\}} \Bigl\{K(x,t;{\{\xi\}}|x',t';\{\xi'\}\Bigr\}=
 \sqrt{\frac {\Omega_{0}}{\pi }}\int
 K(x,t;{\{\xi\}}|x',t;\{\xi\})\,D\{\xi\},
\end{equation}
and correspondingly the operation $ Sp_{x}$ is defined as a simple
integration
\begin{equation}
\label{2.07}
 Sp_{x} \Bigl\{K(x,t;{\{\xi\}}|x',t';\{\xi'\}\Bigr\}=
 \sqrt{\frac {\Omega_{0}}{\pi }}\int
 K(x,t;{\{\xi\}}|x,t;\{\xi'\})\,dx.
\end{equation}
If one wishes to have the quantity describing irreversible
behavior of the system, it is necessary to change definition of
entropy.
\begin{defin} The von Neumann entropy, the standard measure of
randomness of statistical ensemble described by density  matrix ,
which is defined as:
\begin{equation}
S_N(\epsilon,t) =-\frac{1}{N(t)}Sp_x \Bigl\{\rho \ln{\rho}\Bigr\},
\label{2.08}
\end{equation}
 where $ N(t)=Sp_x\rho $ and $ \rho=Sp_{\{ \xi\}}\{\rho_{stc}\}.$

The definition (\ref{2.08}) of entropy is correct for quantum
information theory and also is agree with the Shannon entropy in
the classical limit.

Often it is interesting to know entropies of isolated quantum
state (partial entropy)
\begin{equation}
\label{2.09}
 S^{(m)}_N( \epsilon,t) =-\frac{1}{N_{m}(t)}
Sp_x \Bigl\{\rho^{(m)}\ln{\rho^{(m)}}\Bigr\},
\end{equation}
where $N_{m}(t)=Sp_x\rho^{(m)}$ and $ \rho^{(m)}=Sp_{\{ \xi\} }
\Bigl\{\rho^{(m)}_{stc}\Bigr\}$.
\end{defin}

\begin{defin}
 The entropies can be defined another forms. The total
entropy may be calculated by formula
\begin{equation}
\label{2.10}
 S_G(\epsilon,t) =-\frac{1}{N(t)}Sp_x\biggl\{
 Sp_{\{\xi\}}\Bigl[\rho_{stc}\ln{\rho_{stc}}\Bigr]\biggr\},
 \end{equation}
and the partial entropy correspondingly by formula
\begin{equation}
 S_G^{(m)}(\epsilon,t) =-\frac{1}{N_m(t)}Sp_x\biggl
 \{ Sp_{\{\xi\}}\Bigl[\rho_{stc}^{(m)}
\ln{\rho_{stc}^{(m)}}\Bigr]\biggr\}. \label{2.11}
\end{equation}
\end{defin}
Before proceeding further to the calculations of the physical
parameters let us write down the general form of the partial
stochastic density matrix:
$$
\rho _{stc}^{( m)}\Bigl( x,t;\{\xi\} | x^{\prime },t^{\prime
};\{\xi^{\prime }\}\Bigr) =\sqrt{\frac { \Omega_{0}}{\pi
r(t)r(t')}}\exp \biggl\{-i\Bigl(m+\frac 12\Bigr){\Omega_0}
\biggl[\int\limits_{t_{0}}^t \frac{d\mu}{r^2(\mu)}-
\int\limits_{t_{0}}^{t'}\frac{d\mu}{r^2(\mu)}\biggr]+
$$
\begin{equation}
\label{2.12}
 \frac{i}{2}\biggl[ \frac{r_t(t)}{r(t)}x^2-
 \frac{r_{t'}(t')}{r(t')}{x'}^2\biggr]-\frac12\Omega_{0}
 \biggl[\frac{1}{r^2(t)}x^2
 +\frac{1}{r^2(t')}{x'}^2\biggr] \biggr\}H_m\Bigl(\sqrt{\Omega_0}
 \frac{x}{r(t)}\Bigr)H_m\Bigl(\sqrt{\Omega_0}\frac{x'}{r(t')}\Bigr).
\end{equation}

\setcounter{equation}{0}
\section{Calculation of thermodynamic potentials}
\label{sec-3}

Now let us turn to the calculation of the ground state equilibrium
entropy, which is defined by the expression (\ref{2.11}):
\begin{equation}
\label{3.01}
 S^{(0)}_G(\lambda)=\lim\limits_{t\rightarrow +\infty
} S^{(0)}(\lambda,t).
\end{equation}

The solution of equation (\ref{1.08}) may be presented to the
form:

\begin{equation} \xi (t)=  \Biggl\{
\begin{array}{ll}
\xi_0(t)\equiv\exp({i\Omega_{0}t}),&t\leq t_0=-\infty,\\
\xi_0(t_0)\exp{\Bigl \{ \stackrel{t}{\mathrel{\mathop{\int
}\limits_{t_0}}} \Phi(t')dt'\Bigr \}},&t>t_0 ,
\end{array}
\label{3.02}
\end{equation}
where $\Phi(t)$ is some complex function.

 After the substitution (\ref{3.02}) in the (\ref{1.08}) we
can define the following nonlinear  SDE for a function $\Phi (t)$:
\begin{equation}
\label{3.03}
 \dot \Phi+\Phi^2+\Omega_ {0}^2
+\sqrt{2\epsilon}f(t)=0, \quad \Phi (t_0)=\dot
\xi_0(t_0)/\xi_0(t_0)=i\Omega_{0},
\end{equation}
where $ \dot \Phi=d_t\Phi$. The second equation in the
(\ref{3.02}) expresses a condition which guarantees continuity of
the function $ \xi (t) $ and its first derivative at the $t=t_0$.
The function $\Phi (t) $ is described a complex-valued random
process due to the initial condition. As a result the SDE
(\ref{3.03}) is equivalent to a set of two SDE for real-valued
random processes. Namely, introducing real and imaginary parts of
$ \Phi (t) $
$$
\Phi (t) =u_1 (t) +iu_2 (t),
$$
we finally obtain the following set of SDE for the components of
random vector process $ \vec{u}\equiv \vec{u}(u_1,u_2) $:

\begin{equation}
\left \{
\begin{array}{l}
\dot u_1=-u_1^2+u_2^2-\Omega_{0}^2-\sqrt{2\epsilon}f(t),\\
\dot u_2=-2u_1u_2,\\
\end{array}
\right. \ \left \{
\begin{array}{lll}
u_1(t_0)={Re[\dot \xi_0(t_0)/\xi_0(t_0)]}=0,\\
u_2(t_0)=Im[\dot \xi_0(t_0)/\xi_0(t_0)]=\Omega_{0}.\\
\end{array} \right.
\label{3.04}
\end{equation}
The pair of random processes $(u_1,u_2)$ are not independent,
because their evolution is influenced by the common random force $
f (t)$. This means that the joint probability distribution:
$$
P_0(\vec u,t\vert \vec u_0,t_0 )= \biggl <\prod_{i=1}^2\delta
(u_i(t)-u_{0i})\biggr>, \quad u_{0i}=u_i(t_0),
 $$
is a non-factorable function. Proceeding from the known evolution
equations (\ref{3.04}), we obtain by the standard method the
Fokker-Planck equation for $ P_0 $ (see for ex. \cite{Gred} or
\cite{Gard}) which has the form:
\begin{equation}
\label{{3.05}} \frac {\partial P_0}{\partial t}=\hat L_0P_0,
\end{equation}

\begin{equation}
\label{3.06}
 \hat L_0(u_1,u_2)\equiv \epsilon \frac
{\partial^2}{\partial u_1^2}+ (u_1^2-u_2^2+\Omega_{0}^2) \frac
{\partial }{\partial u_1}+2u_1u_2 \frac {\partial }{\partial u_2}+
4u_1,
\end{equation}
with the initial and border conditions:
\begin{equation}
\label{3.07}
 P_0(u_1,u_2;t)\Bigl|_{t=t_0}=\delta(u_1-u_{01})\delta(u_2-u_{02}),\quad
 P_0(u_1,u_2;t)\Bigl|_{||\vec{u}||\rightarrow+\infty}\rightarrow0.
\end{equation}

 Proceeding from
(\ref{2.08}), (\ref{2.09}) and (\ref{2.10}), we can write down the
expression for nonequilibrium entropy of ground state:
\begin{equation}
\label{{3.08}}
 S^{0}_G(\epsilon,t)=-\frac{1}{2}+
\frac{N_{\alpha;\alpha}(t)}{ N_\alpha(t)}\biggl|_{\alpha=0},\quad
 \end{equation}
 where
\begin{equation}
\label{3.09}
 N_{\alpha}(t)=Sp_{\{\xi\}}\Bigl\{I_{\alpha}(t;\{\xi\})\Bigr\},\quad
 N_{\alpha;\alpha}(t)=\partial_{\alpha}Sp_{\{\xi\}}\Bigl\{
 I_{\alpha}(t;\{\xi\})\Bigr\}.
 \end{equation}

In (\ref{3.09}) the following notation have been made:

$$
  I_{\alpha}(t;\{\xi\})=\frac{\theta\bigl(u_2(t)\bigr)}{\sqrt{u_2(t)}}
\exp\biggl(-(\alpha+1)\int\limits_{t_0}^tu_1(t')dt'\biggr),\quad
 \theta(u_2)=  \biggl\{
\begin{array}{ll}
1,&u_2>0,\\
0,&u_2<0 ,
\end{array}
$$
Now we are ready to calculate the functional traces in expression
(\ref{3.09}). Using the distribution $P_0$ it is easy to construct
continuum measures in the expressions for
$N_0(t)=N_{\alpha}(t)|_{\alpha=0}$ and $N_{\alpha;\alpha}(t)$,
which define the entropy of non-equilibrium quantum systems.
 These integrals may
be calculated with the use of the generalized Feynman-Kac formula
(see for example \cite {ash1}). So as give the following
representation:

\begin{equation}
\label{3.10}
 N_{\alpha}(t)=\int\limits_{-\infty}^{+\infty}
\int\limits_{-\infty}^{+\infty}\frac{1}{\sqrt{u_2}}\theta(u_2)
Q_\alpha (u_1,u_2;t)du_1du_2,
\end{equation}
 where the function $Q_\alpha (u_1,u_2,t)$ is a solution of the
equation:

\begin{equation}
\label{3.11}
\partial_t Q_\alpha =\hat{L}_0Q_\alpha
-(\alpha +1)u_1 Q_\alpha,
\end{equation}
which is satisfied the following initial and border conditions:
$$
  Q_\alpha(u_1,u_2;t)\Bigl|_{t=t_0}=\delta(u_1-u_{01})
 \delta(u_2-u_{02}),\quad Q_{\alpha}(u_1,u_2;t)
 \Bigl|_{||\vec{u}||\rightarrow+\infty}\rightarrow 0.
 $$
Remember that if in the expression (\ref{3.10}) the substitution
$\alpha=0$ is put, the normalization constant $N_0(t)$ will be
obtained. If we can calculate the quantity
$Q_\alpha(u_1,u_2;t)$ then obviously we can be able calculate the
function $ D_\alpha (u_1,u_2,t)\equiv
\partial_\alpha Q_\alpha (u_1,u_2,t) $.  It is easy to obtain the
equation for the latter by differentiae the equation (\ref{3.11})
with respect to $\alpha$:

\begin{equation}
\label{{3.12}}
 \partial_t D_\alpha =\hat{L}_0D_\alpha
-(\alpha+1) u_1 D_\alpha-u_1 Q_\alpha,
  \end{equation}
  correspondingly with conditions:
$$
 D_\alpha(u_1,u_2;t)\Bigl|_{t=t_0}=0,\quad D_\alpha(u_1,u_2;t)
  \Bigl|_{||\vec{u}||\rightarrow+\infty}\rightarrow0.
$$
 Introducing the designations
 $ D_0 (u_1,u_2;t)\equiv D_\alpha (u_1,u_2;t)\Bigl|_{\alpha=0}$,
 we obtain the  representation:
\begin{equation}
\label{{3.13} }
 N_{0;0}(t)=N_{\alpha;\alpha}(t)\Bigl|_{\alpha=0}=
 \int\limits_{-\infty}^{+\infty}
\int\limits_{-\infty}^{+\infty}\frac{1}{\sqrt{u_2}}
\theta(u_2)D_0(u_1,u_2;t)du_1 du_2.
\end{equation}

Now the expression for the equilibrium entropy can be obtained:

\begin{equation}
\label{{3.14}}
  S^{(0)}_G(\lambda)=-\frac{1}{2}+\frac{N_{0;0}^{st}(\lambda)}{N_0^{st}(\lambda)},
\end{equation}
where
$$
 N_0^{st}(\lambda)=\lim\limits_{t\rightarrow\infty}N_{0}(t), \quad
N_{0;0}^{st}(\lambda)=\lim\limits_{t\rightarrow\infty}N_{0;0}(t),\quad
\lambda=(\Omega_0/\epsilon^{1/3})^2.
$$

Remember that at the limit $t\rightarrow+\infty$ the solutions
$Q_\alpha(u_1,u_2;t)$ and $D_0(u_1,u_2;t)$ turn to their
stationary limits $Q_\alpha^{st}(u_1,u_2)$ and $D_0^{st}(u_1,u_2)$
correspondingly.

Note that the von Neumann entropy (\ref{2.08}) coincides with the
entropy expressions in (\ref{2.10}) when the interaction with the
environment is weak, i.e. then $\epsilon\ll 1$.

To have in view the expressions (\ref{2.05}) and (\ref{2.12}) by a
little transformation for the average energy it can be defined:

\begin{equation}
E_{osc}^{(0)}(\lambda
)=\lim\limits_{t\rightarrow+\infty}\biggl\{\frac{1}{N_0(t)}Sp_xSp_{\{
\xi\} }\Bigl[\hat H_0 \rho_{stc}^{(0)}\Bigl]\biggr\} ,
\label{3.15}
\end{equation}

where operator $\hat H_0$ is determined from (\ref{1.02}) by
averaging of $\hat H$ over all:
\begin{equation}
 \hat H_0 = -\frac {1}{2}\frac {\partial^2}{\partial x^2}+
\frac {1}{2}\Omega^2_0x^2. \label{3.16}
 \end{equation}
 Substituting (\ref{3.16}) in (\ref{3.15})
after simple algebra we obtain for the ground state energy:

\begin{equation}
\label{3.17}
 E^{(0)}_{osc}( \lambda) =\frac 12\Bigl(
1+K( \lambda )\Bigr)\Omega _{0} ,
\end{equation}
with the designations:
$$
K( \lambda) =\frac{1}{N_0^{st}(\lambda)}
\int\limits_{-\infty}^{+\infty}\int\limits_{0}^{+\infty}
\frac{1}{\sqrt{\bar{u}_2}}\biggl\{-1+\frac{\bar{u}_1^2+
\bar{u}_2^2+\lambda}{2\sqrt{\lambda}\bar{u}_2}\biggr\}
Q_0^{st}(\lambda;\bar{u}_1,\bar{u}_2)\,d\bar{u}_1\,d\bar{u}_2,
$$
\begin{equation}
\label{3.18}
N_0^{st}(\lambda)=\int\limits_{-\infty}^{+\infty}\int\limits_{0}^{+\infty}
\frac{1}{\sqrt{\bar{u}_2}}Q_0^{st}(\lambda;\bar{u}_1,\bar{u}_2)\,d\bar{u}_1\,d\bar{u}_2,
\quad \bar{u}_1=\frac{u_1}{\epsilon^{1/3}},\quad
\bar{u}_2=\frac{u_2}{\epsilon^{1/3}}.
\end{equation}
In the expression (\ref{3.18}) the function
$Q_{0}(\bar{u}_1,\bar{u}_2;t)$ is a solution of the equation:
\begin{equation}
\frac{\partial{Q_0(\bar{u}_1,\bar{u}_2;t)}}{\partial{t}}=\biggl\{\frac
{\partial^2}{\partial \bar{u}_1^2}+
(\bar{u}_1^2-\bar{u}_2^2+\lambda) \frac {\partial }{\partial
\bar{u}_1}+2\bar{u}_1\bar{u}_2 \frac {\partial }{\partial\bar{
u}_2}+ 3\bar{u}_1\biggr\}Q_0(\bar{u}_1,\bar{u}_2;t), \label{3.19}
\end{equation}
in the limit of stationary possesses
$Q_0^{st}(\lambda;\bar{u}_1,\bar{u}_2)=\lim_{t\rightarrow+\infty}
 Q_0(\bar{u}_1,\bar{u}_2;t)$.

 The energy of $n$-th quantum state in the thermodynamic
 limit is calculated similarly:

 \begin{equation}
E^{(n)}_{osc}( \lambda) =\bigl(n+1/2\bigr)\bigl(1+K( \lambda
)\bigr)\Omega _{0}.
 \label{3.20}
\end{equation}

As evident from expression (\ref{3.20}) after the relaxation all
energetic levels are equidistant.

\setcounter{equation}{0}
\section{Uncertain relations, Weyl transformation and
 Wigner function for the ground state}
\label{sec-4}

According to the uncertainty relation, in quantum system the
coordinates and momentums can't have arbitrary small dispersions.
This principle is experimentally verified many times. However at
the present time as a result of recent quantum technology
development the necessity in the overcoming of this fundamental
restriction and taking the control over the uncertainty relation
arises.

The dispersion of the operator $\hat{A}_i$ is determined by:

\begin{equation}
\Delta\hat{A}_i(t)\equiv\biggl\{Sp_x\Bigl(\rho\hat{A}
_i^2\Bigl)-\Bigl[Sp_x\Bigl(\rho\hat{A} _i
\Bigr)\Bigr]^2\biggr\}^{1/2}.
 \label{4.01}
\end{equation}
Now using the expression (\ref{4.01}) it is easy to calculate the
dispersions for the operator $\hat{A}_i$ in  the extended space
$\Xi=R^1\otimes {R_{\{\xi\}}}$ at the time point $t$:

\begin{equation}
\Delta\hat{A}_i(t)\equiv\frac{1}{N_0(t)}\biggl\{Sp_xSp_{\{\xi\}}
\Bigl(\rho_{stc}^{(0)}\hat{A}
_i^2\Bigl)-\Bigl[Sp_xSp_{\{\xi\}}\Bigl(\rho_{stc}^{(0)}\hat{A} _i
\Bigr)\Bigr]^2\biggr\}^{1/2}.
 \label{4.02}
\end{equation}
Now, with help of expression (\ref{4.02}) we can calculate the
dispersions for the coordinate $\hat{x}$ and momentum $\hat{p}$ :
\begin{equation}
\Delta\hat{x}(t)=\biggl\{\frac{1}{2N_0(t)}\int\limits_{-\infty}^{+\infty}
\int\limits_{0}^{+\infty}
\frac{1}{{\bar{u}_2}^{3/2}}Q_{0}(\bar{u}_1,\bar{u}_2;t)
\,d\bar{u}_1\,d\bar{u}_2\biggr\}^{1/2}, \label{4.03}
\end{equation}
\begin{equation}
\Delta\hat{p}(t)=\biggl\{\frac{1}{2N_0(t)}\int\limits_{-\infty}^{+\infty}
\int\limits_{0}^{+\infty}
\frac{\bar{u}_1^2+\bar{u}_2^2}{{\bar{u}_2}^{3/2}}Q_0(\bar{u}_1,\bar{u}_2;t)
\,d\bar{u}_1\,d\bar{u}_2\biggr\}^{1/2}.
 \label{4.04}
\end{equation}

The product of dispersions for the operators at time point
 $t_0$, when the interaction with the environment is switch on
 describes the standard Hisenberg relation
$[\Delta\hat{x}(t)\Delta\hat{p}(t)]\Bigl|_{t=t_0}=1/2$. It is
interesting to calculate the uncertainty relation for large time
values in the equilibrium limit. By averaging $t\rightarrow\infty$
we obtain:

\begin{equation}
\Delta\hat{x}_{st}\Delta\hat{p}_{st}=\lim\limits_{t\rightarrow+\infty}
\Bigl[\Delta\hat{x}(t)\Delta\hat{p}(t)\Bigr]=
\frac{1}{2}\frac{\sqrt{A_x(\lambda)A_p(\lambda)}}{N_0^{st}(\lambda)},
 \label{4.05}
\end{equation}
where the following notations have been made:

$$
A_x(\lambda)=\int\limits_{-\infty}^{+\infty}
\int\limits_{0}^{+\infty}
\frac{1}{{\bar{u}_2}^{3/2}}Q_0^{st}(\lambda;\bar{u}_1,\bar{u}_2)
\,d\bar{u}_1\,d\bar{u}_2,
$$
\begin{equation}
A_p(\lambda)=\int\limits_{-\infty}^{+\infty}
\int\limits_{0}^{+\infty}
\frac{\bar{u}_1^2+\bar{u}_2^2}{{\bar{u}_2}^{3/2}}Q_0^{st}
(\lambda;\bar{u}_1,\bar{u}_2)\,d\bar{u}_1\,d\bar{u}_2.
\label{4.06}
\end{equation}
where $Q^{st}_0(\lambda;\bar{u}_1,\bar{u}_2)$ is the stationary
limit of solution of equation (\ref{3.11}) at $\alpha=0$.

It is easy to check that the relations (\ref{4.05})-(\ref{4.06})
differ substantially from the Hisenberg uncertainty relations. In
particular, it allows to control the fundamental relation
(\ref{4.05}) using the power parameter $\lambda$, which
characterizes the fluctuations of the environment.

\begin{defin}
We call the expression
\begin{equation}
\label{4.07}
 W_{stc}( p,x,t;\{\xi\}) = \sum_{m=0}^\infty w_0^{( m)}W_{stc}
 (m|p,x,t;\{ \xi\}),
\end{equation}
 stochastic Winger function and correspondingly  $W_{stc}(m|p,x,t;\{ \xi\})$ the partial
 stochastic Winger function. In particular, for the partial stochastic Winger function we get:

\begin{equation}
\label{4.08}
 W_{stc}(m|p,x,t;\{\xi\}) = \int\limits_{-\infty}^{+\infty}
 e^{ipv}\Psi_{stc}(m| (x-v/2),t;\{\xi\})
 \Psi_{stc}^{\ast}(m|(x+v/2),t;\{ \xi\})\,dv,
\end{equation}
\end{defin}

Using the stochastic Winger function it is possible to calculate
the mean values of the physical quantities, which correspond to
the operators $\hat{A}$:
\begin{equation}
\label{4.09}
 \bar{A}=\bar{a}\equiv\int\limits_{-\infty}^{+\infty}\int\limits_{-\infty}^{+\infty}
 Sp_{\{\xi\}}\Bigl\{a(p,x,t;\{\xi\})\rho_{stc}^{W}(p,x,t;\{\xi\})\Bigr\}\,dp\,dx,
\end{equation}

where stochastic function $a(p,x,t;\{\xi\})$ is defined by Weyl
transformation of operator $\hat{A}$:
\begin{equation}
\label{4.10}
a(p,x,t;\{\xi\})=\int\limits_{-\infty}^{+\infty}e^{ipv}\Psi_{stc}(m|
(x-v/2),t; \{\xi\})\hat{A}\Psi_{stc}^{\ast}(m| (x+v/2),t; \{
\xi\})\,dv.
\end{equation}

Note that the Weyl transformation of quantum operator $\hat{A}$
after averaging over the random process
$Sp_{\{\xi\}}\{a(p,x,t;\{\xi\})\}$ can be used in order to obtain
the classical quantity corresponding to the quantum operator
$\hat{A}$. This function is the classical limit
$\hbar\rightarrow0$ (the system of units $\hbar=1$ and $c=1$ is
used, so the limit $\hbar\rightarrow 0$ actually corresponds to
infinite action limit):

$$
\hat{A} \xrightarrow{\text{cl}}
a_{cl}(p,x,t)=\lim_{\hbar\rightarrow 0
}Sp_{\{\xi\}}\Bigl\{a(p,x,t;\{\xi\})\Bigr\}.
$$

Now we can construct the Winger function for the ground state:
$$
W^{0}(x,p;t)=Sp_{\{\xi\}}\Bigl\{W_{stc}(0|p,x,t;\{\xi\})\Bigr\}
=
$$
\begin{equation}
\label{4.11}
2\sqrt{\frac{\lambda}{\pi}}\int\limits_{-\infty}^{+\infty}
\int\limits_{0}^{+\infty}
\frac{1}{\sqrt{\bar{u}_2}}\exp\biggl\{-\frac{(p-\bar{u}_1x)^2-\bar{u}_2^2x^2}{\bar{u}_2}
\biggr\}Q(\bar{u}_1,\bar{u}_2;t)\,d\bar{u}_1d\bar{u}_2.
\end{equation}
It is easy to see that function (\ref{4.11}) make sense coordinate
$x$ and momentum $p$ simultaneous distribution  in the phase space
at the time $t$. Particularly with the help of this expression
 may be investigated the relaxation processes and
average distribution in the limit of stationary processes. Note
that such as in the regular case the integration of function
$W_{stc}(n|p,x,t;\{\xi\})$ by phase space is normalized per unit:

\begin{equation}
\label{4.12}
\int\limits_{-\infty}^{+\infty}\int\limits_{-\infty}^{+\infty}
 W_{stc}(n|p,x,t;\{\xi\})\,dpdx=1.
\end{equation}
Finally it is important to note that for averaging function
(\ref{4.11}) it is not to hold expression type of (\ref{4.12}).


\section*{Conclusion}

There are three different reasons which may cause a chaos in the
basic quantum mechanical object, i.e. the wave function. The first
reason refers to measurements performed over a quantum system
\cite{gp,sinai}. The second reason consists in the more
fundamental openness of any quantum system resulting from the fact
that all the beings are immersed into a physical vacuum
\cite{glimm}. In the third place, as it follows from the recent
papers \cite{avt1,avt2,avt3}, a chaos may also appear in the wave
function even in a closed dynamical system. As it is shown in
\cite{sinai}, there is a close connection between a classical
nonintegrability and a chaos in the corresponding quantum system.
Many of the fundamentally important questions of the quantum
physics such as the Lamb shift of energy levels, spontaneous
transitions between the atom levels, quantum Zeno effect
\cite{itano}, processes of chaos and self-organization in quantum
systems, especially those where the phenomena of phase transitions
type may occur, can be described qualitatively and quantitatively
in a rigorous way only within the nonperturbative approaches. The
Lindblad representation \cite{gorini,lindblad} for the density
matrix of the system "quantum object + thermostat" describes {\em
a priori} the most general situation which may appear in the
nonrelativistic quantum mechanics. Nevertheless, we need to
consider a reduced density matrix on a semi-group \cite{gp}, when
investigating a quantum subsystem. This is quite an ambiguous
procedure and moreover its technical realization is possible only
in the framework of a particular perturbative scheme.

A crucially new approach to constructing the quantum mechanics of
the closed nonrelativistic system "quantum object + thermostat" has
been developed recently by the authors of \cite{avt1,avt2} from the
principle of "local correctness of Schr\"{o}dinger representation".
To put it differently, it has been assumed that the evolution of
the quantum system is such that it may be described by the
Schr\"{o}dinger equation on any small time interval, while the
motion as a whole is described by a SDE for the wave function. In
this case, however, there emerges not a simple problem to find a
measure for calculating the average values of the physical system
parameters. Nevertheless, there exists a certain class of models
for which all the derivations can be made not applying the
perturbation theory \cite{avt2}.

In the present paper we explore further the possibility of
building the nonrelativistic quantum mechanics of closed system
"quantum object + thermostat" within the framework of the model of
one-dimensional randomly wandering QHO (with a random frequency
and subjected to a random external force). Mathematically the
problem is formulated in terms of SDE for a complex-valued
probability process defined on the extended space $ R^1\otimes
R_{\{ \xi \} }. $ The initial SDE is reduced to the
Schr\"{o}dinger equation for an autonomous oscillator defined on a
random space-time continuum, with the use of a nonlinear
transformation and one-dimensional etalon nonlinear equation of
the Langevin type defined on the functional space $ R_{\{ \xi\} }.
$ It is possible to find for any fixed $ \{ \xi\} $ an orthonormal
basis of complex-valued random functionals in the space $ L_2(R^1)
$ of square-integrable functions. With the assumption that the
random force generator is described by a white noise correlator,
the Fokker-Planck equation for a conditional probability is found.
From the solutions of this equation on an infinitely small time
interval a measure of the functional space $ R_{\{ \xi \} }$ can
be constructed. Then by averaging an instantaneous value of the
transition probability over the space $ R_{\{\xi \} } $, the mean
value of the transition probability is represented by a functional
integral. Using the generalized Feynman-Kac theorem, it is
possible to reduce the functional integral in the most general
case, where both frequency and force are random, to a multiple
integral of the fundamental solution of some parabolic partial
differential equation. The qualitative analysis of the parabolic
equation shows that it may have discontinuous solutions. This is
equivalent to the existence of phenomena like the phase
transitions in the microscopic transition probabilities. In the
context of the developed approach the representation of the
stochastic density matrix is introduced, which allows to build a
closed scheme for both nonequilibrium and equilibrium
thermodynamics. The analytic formulas for the ground energy level
broadening and shift are obtained, as well as for the entropy of
the ground quantum state. The important results of the work are
calculation of expressions for uncertain relations and Wigner
function for the quantum subsystem which is strong interacting
with the environment.

The further development of the considered formalism in application
to exactly solvable many-dimensional models may essentially extend
our understanding of the quantum world and lead us to the new
nontrivial discoveries.


\end{document}